\documentclass[twocolumn,prl,showpacs,amsmath,assymb,eufrak]{revtex4}
\usepackage{epsfig,amsfonts,amsbsy}
\newcommand{\ret}{\nonumber\\}

\newcommand{\sbkt}[1]{\langle#1\rangle}
\newcommand{\bbkt}[1]{\bigl\langle#1\bigr\rangle}
\newcommand{\Bbkt}[1]{\Bigl\langle#1\Bigr\rangle}

\newcommand{\Ttot}{\mathcal{T}}
\newcommand{\WT}{\mathcal{W}}
\newcommand{\calD}{\mathcal{D}}
\newcommand{\pathG}{\hat{\Gamma}}
\newcommand{\patha}{\hat{\alpha}}
\newcommand{\pathGd}{\hat{\Gamma}^\dagger}
\newcommand{\pathad}{\hat{\alpha}^\dagger}
\newcommand{\thetae}{\theta^\mathrm{ex}}
\newcommand{\Thetae}{\Theta^\mathrm{ex}}

\newcommand{\Jex}{J^\mathrm{ex}}
\newcommand{\rhost}{\rho^\mathrm{st}}

\newcommand{\oep}{O(\epsilon^3)}
\newcommand{\oet}{O(\epsilon^2)}

\newcommand{\Di}{\mathit{\Delta}}

\newcommand{\pa}{\nu}
\newcommand{\Gi}{\Gamma_\mathrm{i}}
\newcommand{\Gf}{\Gamma_\mathrm{f}}
\newcommand{\ai}{\alpha_\mathrm{i}}
\newcommand{\af}{\alpha_\mathrm{f}}

\newcommand{\intai}{\int_0^\Ttot dt}
\newcommand{\ste}{\mathrm{st}}
\newcommand{\sumk}{\sum_{k=1}^2}

\newcommand{\Dep}{\Delta}
\begin{document}
\title{Steady State Thermodynamics for Heat Conduction --- Microscopic Derivation}
\author{Teruhisa S. Komatsu${}^1$, Naoko Nakagawa${}^2$, Shin-ichi Sasa${}^3$, and 
Hal Tasaki${}^4$}
\affiliation {${}^{1,3}$
Department of Pure and Applied Sciences, The University of Tokyo,
 Komaba, Meguro-ku, Tokyo 153-8902, Japan
 \\
 ${}^2$College of Science, 
 Ibaraki University, Mito, Ibaraki 310-8512, Japan
\\
${}^4$Department of Physics, Gakushuin University, Mejiro, Toshima-ku, Tokyo 171-8588, Japan}

\date{\today}

\begin{abstract}
Starting from microscopic mechanics, 
we derive thermodynamic relations for 
heat conducting nonequilibrium steady states.
The extended Clausius relation enables one to 
experimentally determine nonequilibrium entropy 
to the second order in the heat current.
The associated Shannon-like microscopic 
expression of the entropy is suggestive.
When the heat current is fixed, the extended Gibbs 
relation provides a unified treatment of thermodynamic 
forces in linear nonequilibrium regime.
\end{abstract}


\maketitle

Thermodynamics (TD) is a theoretical framework that describes universal
quantitative laws obeyed by  macroscopic systems in equilibrium.
A core of TD is the Clausius relation  $\Di S=Q/T$, which relates the entropy with the heat transfer caused by a change in the system.
Combined with the energy conservation, the Clausius relation leads to the Gibbs relation  $T\,dS=dU+\sum_if_id\nu_i$, where $\nu_i$ is a controllable parameter and $f_i$ the corresponding generalized force.
The Gibbs relation is particularly useful since it represents the  forces as gradients of suitable thermodynamic potentials.
It also played a key role when Gibbs constructed 
equilibrium statistical mechanics.

Here we wish to address the fundamental question whether
TD can be extended to nonequilibrium steady states (NESS) which, like equilibrium states, lack macroscopic time-dependence.
We shall call the possible extension Steady State Thermodynamics (SST).
The possibility of SST is far from trivial since NESS exhibit many properties which are very different from equilibrium states.
First of all a naive extension of the Clausius relation to NESS is never possible since the heat transfer $Q$ generally diverges linearly in time.
It is also a deep theoretical question whether the long range correlation universally observed in NESS \cite{LRC} is consistent with SST.
In addition to such abstract interests, there are nonequilibrium phenomena which may be better understood using SST.
An interesting example is the force exerted on a small rigid body placed in a heat conducting fluid \cite{Wiegand}. 
This force may be 
understood as a thermodynamic force in SST (see \cite{BDB} for related ideas).

NESS sufficiently close to  equilibrium can be characterized by the  linear response theory. 
But this theory, which requires an ensemble of trajectories in space-time, does not lead us directly to SST.
It is also clear that the theory only gives the result up to the first order in the ``degree of nonequilibrium.''

Although an extension of TD to NESS (or, equivalently, a construction of SST) may sound as a formidably difficult task, there are at least two branches of studies which are  encouraging.
One is the series of works which reveal deep implications on NESS of the microscopic time-reversal symmetry.
It has been shown that the simple symmetry \eqref{e:symmetry}
leads to various nontrivial 
results including the Green-Kubo relation, 
Kawasaki's non-linear response relation, 
and the fluctuation theorem  \cite{symmetry2,symmetry1}.
Although none of these works directly treat  extensions of  TD,  techniques for characterizing NESS and energy transfer may be useful.
The other is a series of works in which theoretical consistency of SST was examined from purely phenomenological points of view.
These works provide us with some useful guidelines for constructing SST.
In \cite{OP} (see also \cite{La}) it was proposed that heat $Q$ in the Clausius relation should be replaced by ``excess heat'', which is the intrinsic heat transfer caused by the change of the state.
In \cite{ST} it was conjectured that one should fix the total heat current $J$ to get a Gibbs relation in a heat conducting NESS.

In the present Letter, we shall report a unification of these two  branches, which has led us to a microscopic  construction of  SST
(see \cite{HS,ST} for early attempts to construct SST).
More precisely we start from  microscopic mechanics, 
and derive a  natural extension of the Clausius relation
to heat conducting NESS.
The extended Clausius relation enables one to experimentally 
determine  nonequilibrium entropy to the second order in the 
heat current. 
This extends the construction of entropy for NESS by Ruelle \cite{Ruelle}, who treated a simpler
system with isokinetic thermostat.
We further determine the precise form \eqref{e:Ssym} of the entropy.
In systems with a fixed heat current, 
we derive an extension of the Gibbs relation, which enables 
one to treat thermodynamic forces in linear nonequilibrium 
regime in a  new unified manner.


\paragraph*{Setup:}
Our theory can be developed in various nonequilibrium settings 
including driven or sheared fluid. For simplicity we here focus 
on heat conduction, and consider a system which is attached to two heat baths and has  
controllable  parameters (such as the volume).

We assume that the system consists of $N$ particles  
whose coordinates are collectively denoted as 
$\Gamma=(\mathbf{r}_1,\ldots,\mathbf{r}_N;\mathbf{p}_1,\ldots,\mathbf{p}_N)$.
We write  its time-reversal as
$\Gamma^*
=(\mathbf{r}_1,\ldots,\mathbf{r}_N;-\mathbf{p}_1,\ldots,-\mathbf{p}_N)$.
When discussing time evolution of $\Gamma$, we denote by $\Gamma(t)$ 
its value at time $t$, and by $\pathG=(\Gamma(t))_{t\in[0,\Ttot]}$ its history 
(or path) over the time interval $[0,\Ttot]$. Given a path $\pathG$, 
we denote its time reversal as $\pathGd=((\Gamma(\Ttot-t))^*)_{t\in[0,\Ttot]}$.

We take a Hamiltonian satisfying the time-reversal 
symmetry $H_\pa(\Gamma)=H_\pa(\Gamma^*)$, where $\pa$ is the set of controllable 
parameters. Time evolution of the system is determined by the Hamiltonian and 
coupling to the two external heat baths with inverse temperatures $\beta_1$ and
$\beta_2$. To model the heat baths, one may use  Langevin noise or an explicit 
construction using Hamiltonian mechanics as in \cite{KNST}. Our results are 
valid in both (and other physically natural) settings. We shall characterize 
our system using the set of parameters $\alpha=(\beta_1,\beta_2;\pa)$.

An external agent performs an operation to the system 
by changing $\alpha$ according to a prefixed protocol. A protocol is 
specified by a function $\alpha(t)=(\beta_1(t),\beta_2(t);\pa(t))$ of 
$t\in[0,\Ttot]$. We denote by $\patha=(\alpha(t))_{t\in[0,\Ttot]}$ 
the whole protocol. Again $\pathad=(\alpha(\Ttot-t))_{t\in[0,\Ttot]}$ 
denotes the time-reversal of $\patha$. By $(\alpha)$ we denote a protocol 
in which the parameters are kept constant at $\alpha$.

Consider a time evolution with a protocol $\patha$, and denote the 
probability weight for a path $\pathG$ as $\WT_{\patha}[\pathG]$.
It is normalized as $\int_{\Gamma(0)=\Gi}\calD\pathG\,\WT_{\patha}[\pathG]=1$ 
for any initial state $\Gi$, where $\calD\pathG$ denotes the path integral 
over all the histories (with the specified initial condition).

 
\paragraph*{Time-reversal symmetry:}
Let $J_k(\pathG;t)$ be the heat current from the $k$-th bath to the system at time $t$ in the history $\pathG$. 
Then the entropy production rate at $t$ is given by
$\theta_{\patha}(\pathG;t)=-\sumk\beta_k(t)\,J_k(\pathG;t)$.
By integration, we get the entropy production $\Theta_{\patha}(\pathG)=\intai\,\theta_{\patha}(\pathG;t)$.

Then it has been shown \cite{symmetry1,KNST}  that the present (and other physically realistic) time evolution
satisfies
\begin{equation}
\WT_{\patha}[\pathG]=\WT_{\pathad}[\pathGd]\,e^{\Theta_{\patha}(\pathG)},
\label{e:symmetry}
\end{equation}
which is the basis of the present work.

\paragraph*{Steady state and its representation:}
We assume that the system settles to a unique NESS when it evolves for a sufficiently long time with fixed $\alpha$.
We take $\Ttot$  much larger than the relaxation time.
We treat NESS with a small heat current, where convection hardly takes place.

In the NESS characterized by $\alpha$, 
the expectation value of  the current  $J_k(\pathG;t)$ takes a constant value, 
which we denote as $\bar{J}_k(\alpha)$. 
We define the excess heat current from the $k$-th bath as 
$\Jex_{k,\patha}(\pathG;t)=J_k(\pathG;t)-\bar{J}_k(\alpha(t))$.
Then $\thetae_{\patha}(\pathG;t)=-\sumk\beta_k(t)\,\Jex_{k,\patha}(\pathG;t)$ 
and its integration $\Thetae_{\patha}(\pathG)=\intai\,\thetae_{\patha}(\pathG;t)$ are  the excess entropy production rate and the  excess entropy production, respectively.

We denote by  $\rhost_{\alpha}(\Gamma)$  the probability distribution for 
the NESS  characterized by $\alpha$.
By using \eqref{e:symmetry}, it was shown in \cite{KN} (see also \cite{KNST}) that the 
distribution has a concise representation
\begin{equation}
\rhost_{\alpha}(\Gamma)
=\exp\Bigl[-S(\alpha)+\frac{\bbkt{\Thetae_{(\alpha)}}^{(\alpha)}_{\ste,\Gamma}
-\bbkt{\Thetae_{(\alpha)}}^{(\alpha)}_{\Gamma^*,\ste}}{2}+\tilde{R}(\alpha,\Gamma)\Bigr],
\label{e:KNrep}
\end{equation}
where $S(\alpha)$ is determined by normalization, and $\tilde{R}(\alpha,\Gamma)=\oep$.
Here the  ``degree of nonequilibrium'' $\epsilon$ is a dimensionless 
quantity proportional to the typical heat current. 
Throughout the present Letter, $\sbkt{\cdots}^{\patha}_{\Gi,\Gf}$ 
stands for the expectation taken 
with respect to the path probability $\WT_{\patha}[\pathG]$ with the 
initial and the final conditions $\Gi$ and $\Gf$, respectively, 
where ``$\ste$'' denotes the steady state \cite{en:average}.
The representation \eqref{e:KNrep} plays a fundamental role in our construction of SST.



\paragraph*{Extended Clausius relation:}

Our first result is a natural extension of the Clausius relation.

Let $\patha$ be an arbitrary quasi-static protocol 
in which the parameters change slowly and 
smoothly from $\ai=\alpha(0)$ to $\af=\alpha(\Ttot)$.
Then   the  extended Clausius relation is
\begin{equation}
S(\af)-S(\ai)=-\bbkt{\Thetae_{\patha}}^{\patha}+R(\patha),
\label{e:Smain2}
\end{equation}
where $R(\patha)$ is a small error about which we discuss shortly.
(Here, and in what follows, 
$\sbkt{\cdots}^{\patha}$ 
is a shorthand for $\sbkt{\cdots}_{\ste,\ste}^{\patha}$.)
Eq. \eqref{e:Smain2} is the core of our SST.

When $\beta_1=\beta_2$, we can show $R(\patha)=0$, and hence \eqref{e:Smain2} becomes 
precisely the standard Clausius relation.
%
Note that the heat current in the original relation has been replaced in the extended relation  \eqref{e:Smain2}  by the excess heat current, 
following the phenomenological proposals  in \cite{OP,La}.
Although the excess entropy production $\sbkt{\Thetae_{\patha}}^{\patha}$ appears 
to depend on paths (in the parameter space) defined by the 
protocol $\patha$, \eqref{e:Smain2} shows, rather strikingly, 
that it can be written as the difference  of the entropy $S(\alpha)$, 
which is a function of $\alpha$.
This is far from a mere consequence of definitions, 
and represents a deep fact that NESS 
possess a nontrivial thermodynamic structure.

For an infinitesimal protocol $\patha$ \cite{en:step}, we will show 
that $R(\patha)=O(\epsilon^2\,\Dep)$, 
where $\Dep$ is a dimensionless quantity which characterizes 
the change $\af-\ai$ \cite{en:Delta}.
 (We know from examples \cite{GCP} that this error estimate is optimal.)
 
The error term $R(\patha)$ for a general quasi-static protocol 
$\patha$ can be obtained by summing up the errors in infinitesimal steps.
In general $O(\Dep)$ sums up to $O(1)$, thus giving $R(\patha)=\oet$.
There are, however, important cases where we can set $R(\patha)=\oep$. 
In such a case, the extended Clausius relation \eqref{e:Smain2} is correct to $\oet$, and hence goes beyond  the linear response theory.
Take, for example, the initial state $\ai$ 
as an equilibrium state with $\beta_1=\beta_2$.
If  we fix $\beta_1$ and change only $\beta_2$, the error $O(\epsilon^2\,\Dep)$ sums up  to $R(\patha)=\oep$ (see also \cite{Ruelle}).

\paragraph*{Nonequilibrium entropy:}
$S(\alpha)$ in \eqref{e:Smain2} was introduced as  the normalization 
factor  in the representation \eqref{e:KNrep}.
It is interesting that this quantity plays the role of entropy in an operational thermodynamic  relation.

In \cite{next}, we shall show 
that this entropy has an interesting symmetrized   Shannon-like expression
\begin{equation}
S(\alpha)=-\int d\Gamma\,\rhost_\alpha(\Gamma)\,
\log\sqrt{\rhost_\alpha(\Gamma)\,\rhost_\alpha(\Gamma^*)}.
\label{e:Ssym}
\end{equation}
Note that the right-hand side becomes precisely the Shannon entropy 
if $\rhost_\alpha(\Gamma)=\rhost_\alpha(\Gamma^*)$. Since the equilibrium 
distribution has this symmetry, 
the entropy $S(\alpha)$ approaches the Shannon entropy in the equilibrium limit.

The expression \eqref{e:Ssym} shows that $S(\alpha)$ reflects certain  
properties of the steady state distribution $\rhost_\alpha(\Gamma)$.
Of particular interest is the long-range correlation \cite{LRC}, 
which should manifest itself as an anomalous size dependence of 
$S(\alpha)$ in the second order in $\epsilon$.
As we have examined above, the extended Clausius relation \eqref{e:Smain2} allows 
one to compare the nonequilibrium and the equilibrium entropies, and
determine $S(\alpha)$  in NESS with  
precision of $\oet$.
One can thus detect the long-range correlation experimentally 
by means of calorimetry.

\paragraph*{Extended Gibbs relation:}

Our second major result is  an  extension of the Gibbs relation.

Let $\beta$ be an arbitrary reference.
Using the energy conservation, and noting that  $\sumk\bar{J}_k(\alpha)=0$, we get
\begin{equation}
\Thetae_{\patha}(\pathG)=\beta\{W_{\patha}(\pathG)+H_{\pa(0)}(\Gamma(0))-H_{\pa(\Ttot)}(\Gamma(\Ttot))\}+\Phi_{\patha}(\pathG),
\label{e:EC}
\end{equation}
where $W_{\patha}(\pathG)$ is the total work done 
by the external agent who changes the parameters $\pa$ (the temperatures of the baths are changed without doing any work), and we defined
$\Phi_{\patha}(\pathG)=-\sumk\intai(\beta_k(t)-\beta)\,\Jex_{k,\patha}(\pathG;t)$.

If the average $\sbkt{\Phi_{\patha}}^{\patha}$ happens to be negligible, then \eqref{e:EC} and the extended Clausius relation \eqref{e:Smain2} imply
\begin{equation}
dS=\frac{dU}{T}+\sum_i\frac{f_i}{T}\,d\pa_i+O(\epsilon^2\,\Dep),
\label{e:dS}
\end{equation}
where we wrote $\sbkt{W_{\patha}}^{\patha}
=-\sum_if_i(\alpha)\,d\pa_i$, with $\pa=(\pa_1,\pa_2,\ldots)$, and 
$f_i(\alpha)$ being the (generalized)  force conjugate to  $\pa_i$.
Remarkably, \eqref{e:dS} is  identical to the standard Gibbs relation.
We stress that all the terms in  \eqref{e:dS} can be determined 
experimentally  by measuring heat currents and mechanical forces.

There may be several ways to make $\sbkt{\Phi_{\patha}}^{\patha}$ negligible.
A natural strategy that comes from the phenomenological proposal in \cite{ST} is to consider a system with a fixed heat current $J$.
To be more precise, we 
consider a ``source-drain system'', in which the two baths  have  different 
characters. The bath 1, which has a lower temperature and coupled 
efficiently to the system, is a ``heat drain.'' It helps the system
to get rid of extra energy  and reach the NESS rapidly. The bath 2, 
which has a higher fixed temperature, is a ``heat source.'' It  supplies 
a constant heat current $J$ to the system in average when the system 
is disturbed by an external operation \cite{en:SD}.
This means that $\sbkt{\Jex_{2,\patha}(t)}^{\patha}$ is negligible.
We now choose the reference as $\beta=\beta_1(0)$ so that $\beta_1(t)-\beta=O(\Dep)$.
Since $\sbkt{\Jex_{1,\patha}(t)}^{\patha}=O(\Dep)$, we find that $\sbkt{\Phi_{\patha}}^{\patha}$ is $O(\Dep^2)$ and hence negligible.

In a source-drain system, it is  natural  to characterize the NESS by parameters 
$(T,J,\pa)$, where $T=1/\beta_1$.  ($\beta_2$ is determined 
from $T$, $J$, and $\nu$.) 
If we restrict ourselves to the operations in which only the parameter 
$\pa$ of the Hamiltonian changes, \eqref{e:dS} gives
\begin{equation}
f_i(T,J,\pa)=-\frac{\partial}{\partial\pa_i}F(T,J,\pa)+O(\epsilon^2),
\label{e:dF}
\end{equation}
where the nonequilibrium free energy is defined by the familiar relation 
$F=U-TS$. The relation \eqref{e:dF} shows that any thermodynamic force 
(including that exerted on a body in a heat conducting fluid)
in the linear nonequilibrium regime is indeed a conservative force 
with the potential $F(T,J,\nu)$.
Although any physical quantity can be  evaluated to $O(\epsilon)$ 
by using the linear response theory, \eqref{e:dF} may provide a novel 
point of view for analyzing thermodynamic forces in the setting 
with a fixed current. For example, 
\eqref{e:dF} implies the Maxwell relation 
$\partial f_i/\partial\pa_j=\partial f_j/\partial\pa_i+\oet$, 
which may be confirmed experimentally in suitable settings.

\paragraph*{Derivation of  main equality \eqref{e:Smain2}:}

We consider an infinitesimal protocol $\patha$  \cite{en:step}.
Noting that $\Theta_{\pathad}(\pathGd)=-\Theta_{\patha}(\pathG)$, \eqref{e:symmetry} implies $\WT_{\patha}[\pathG]\,e^{-\Thetae_{\patha}(\pathG)/2}=\WT_{\pathad}[\pathGd]\,e^{-\Thetae_{\pathad}(\pathGd)/2}$.
By integrating  over all  paths satisfying $\Gamma(0)=\Gi$, $\Gamma(\Ttot)=\Gf$, we get
\begin{equation}
\rhost_{\af}(\Gf)\,\Bbkt{\exp[-\Thetae_{\patha}/2]}^{\patha}_{\Gi,\Gf}
=
\rhost_{\ai}(\Gi^*)\,\Bbkt{\exp[-\Thetae_{\pathad}/2]}^{\pathad}_{\Gf^*,\Gi^*}
\label{e:sym3}
\end{equation}
which is our  starting point.
We later show that
\begin{eqnarray}
&&{
\Bbkt{\exp[-\Thetae_{\patha}/2]}^{\patha}_{\Gi,\Gf}
}\,\,\Bigl/\,{
\Bbkt{\exp[-\Thetae_{\pathad}/2]}^{\pathad}_{\Gf^*,\Gi^*}
}
\ret&&=
\exp\Bigl[-\frac{\bbkt{\Thetae_{\patha}}^{\patha}_{\Gi,\Gf}-
\bbkt{\Thetae_{\pathad}}^{\pathad}_{\Gf^*,\Gi^*}}{2}
+R'(\patha;\Gi,\Gf)\Bigr],
\label{e:eTheta}
\end{eqnarray}
with $R'(\patha;\Gi,\Gf)=\oep+O(\epsilon^2\,\Dep)$.
We here assume that various quantities can be expanded both in $\epsilon$ and $\Dep$.
We regard $\Dep$ as infinitesimal and omit $O(\Dep^2)$.

Note that $\sbkt{\Thetae_{\patha}}^{\patha}_{\Gi,\Gf}=\intai\,\sbkt{\thetae_{\patha}(t)}^{\patha}_{\Gi,\Gf}$ holds, and  
$\sbkt{\thetae_{\patha}(t)}^{\patha}_{\Gi,\Gf}$ attains non-negligible values only when $t$ is near $0$, $\Ttot$, or $\Ttot/2$ (where the system is out of  steady states either by the imposed conditions or the operation).
We can therefore decompose the expectation value as
\begin{equation}
\bbkt{\Thetae_{\patha}}^{\patha}_{\Gi,\Gf}
=
\bbkt{\Thetae_{(\ai)}}^{(\ai)}_{\Gi,\ste}
+
\bbkt{\Thetae_{(\af)}}^{(\af)}_{\ste,\Gf}
+
\bbkt{\Thetae_{\patha}}^{\patha}_{\ste,\ste}.
\label{e:dec}
\end{equation}
By substituting \eqref{e:eTheta} and \eqref{e:dec} into the identity \eqref{e:sym3}, and comparing  the result with the representation \eqref{e:KNrep}, we get
\begin{equation}
S(\af)-S(\ai)=
\frac{1}{2}\Bigl\{\bbkt{\Thetae_{\pathad}}^{\pathad}_{\ste,\ste}-\bbkt{\Thetae_{\patha}}^{\patha}_{\ste,\ste}\Bigr\}+R(\patha),
\label{e:Smain1}
\end{equation}
where $R(\patha)=R'(\patha;\Gi,\Gf)-\tilde{R}(\ai,\Gi)+\tilde{R}(\af,\Gf)$.
Since $R(\patha)=0$ if $\Dep=0$, we must have that $R(\patha)=O(\epsilon^2\,\Dep)$.
Noting the symmetry $\sbkt{\Thetae_{\pathad}}^{\pathad}_{\ste,\ste}=-\sbkt{\Thetae_{\patha}}^{\patha}_{\ste,\ste}$ \cite{next} we get \eqref{e:Smain2} for an infinitesimal process.

\paragraph*{Derivation of \eqref{e:eTheta}:}
We regard (only in this derivation) time-independent
 $H_{\pa(0)}$ as the Hamiltonian of the system,
 and interpret the force from $H_{\pa(t)}-H_{\pa(0)}$ as an ``external force.''
Then the energy balance implies
$H_{\pa(0)}(\Gamma(\Ttot))-H_{\pa(0)}(\Gamma(0))=W(\pathG)+\sumk\intai\,\Jex_k(\pathG;t)$
where $W(\pathG)$ is the total work done by the ``external force''.
By defining 
$\tilde{\Phi}_{\patha}(\pathG)={\Phi}_{\patha}(\pathG)+\beta\,W(\pathG)$,
 we have $\Thetae_{\patha}(\pathG)=\tilde{\Phi}_{\patha}(\pathG)+\beta\{H_{\pa(0)}(\Gamma(0))-H_{\pa(0)}(\Gamma(\Ttot))\}$ as in \eqref{e:EC}.

To simplify notation, we drop $\patha$ or $\pathad$, and abbreviate the expectations $\sbkt{\cdots}^{\patha}_{\Gi,\Gf}$ and $\sbkt{\cdots}^{\pathad}_{\Gf^*,\Gi^*}$ as $\sbkt{\cdots}$ and $\sbkt{\cdots}^\dagger$, respectively.
We make use of the cumulant expansion $\log\sbkt{e^{-\Thetae/2}}=-\sbkt{\Thetae}/2+\sbkt{\Thetae;\Thetae}/8+\cdots$, where $\sbkt{\Thetae;\Thetae}=\sbkt{(\Thetae)^2}-\sbkt{\Thetae}^2$.
Since $H_{\pa(0)}(\Gamma(0))-H_{\pa(0)}(\Gamma(\Ttot))$
 is constant in the present average, we have $\sbkt{\Thetae;\Thetae}=\sbkt{\tilde{\Phi};\tilde{\Phi}}$.
Similar identities  also  hold for  higher order cumulants (see, e.g., \cite{KNST}).

Let us denote by $K$ the left-hand side of \eqref{e:eTheta}.
The cumulant expansion yields
\begin{equation}
\log K=-\frac{\sbkt{\Thetae}-\sbkt{\Thetae}^\dagger}{2}+
\frac{\sbkt{\tilde{\Phi};\tilde{\Phi}}-\sbkt{\tilde{\Phi};\tilde{\Phi}}^\dagger}{8}+O(\tilde{\Phi}^3).
\label{e:logR2}
\end{equation}
To evaluate the second  term, we observe that
\begin{equation}
\sbkt{\tilde{\Phi};\tilde{\Phi}}-\sbkt{\tilde{\Phi};\tilde{\Phi}}^\dagger
=\sbkt{\tilde{\Phi};\tilde{\Phi}}_\mathrm{eq}-\sbkt{\tilde{\Phi};\tilde{\Phi}}_\mathrm{eq}^\dagger+O(\tilde{\Phi}^3),
\label{e:eq}
\end{equation}
where $\sbkt{\cdots}_\mathrm{eq}$ and $\sbkt{\cdots}^\dagger_\mathrm{eq}$ are   averages in the corresponding equilibrium dynamics with the static Hamiltonian $H$ and a common $\beta$.
But the time-reversal symmetry in equilibrium dynamics implies $\sbkt{\tilde{\Phi};\tilde{\Phi}}_\mathrm{eq}=\sbkt{\tilde{\Phi};\tilde{\Phi}}_\mathrm{eq}^\dagger$.
Since $\Phi=O(\epsilon)$ \cite{en:Phi} and $\beta\,W=O(\Dep)$, we have 
$\tilde{\Phi}=O(\epsilon)+O(\Dep)$.
Thus \eqref{e:logR2} and \eqref{e:eq} imply the desired \eqref{e:eTheta}.

\paragraph*{Discussions:}
We treated a general classical model of heat conduction, and derived natural nonequilibrium extensions of the Clausius and the Gibbs relations.
The mere existence of a consistent operational thermodynamics (i.e., SST) may be  of great importance, but the way the extension has been done may also be quite suggestive.

The extended Clausius relation \eqref{e:Smain2} and the associated microscopic expression \eqref{e:Ssym} of the entropy form a theoretical core of the  present work.  They may provide us of a clue to develop a statistical mechanics for NESS that works beyond linear response regime.

It is also suggestive that we obtained the extended Gibbs relation \eqref{e:dS} in a special setting with ``source'' and ``drain'', in which the heat current is fixed.
There is a possibility that this special setting is necessary for uncovering universal statistical properties of heat conducting systems, which properties are hidden in other settings.
In this connection, it is exciting to explore implications of the ``nonequilibrium order parameter'' defined as $\Psi(T,J,\nu)=\partial F(T,J,\nu)/\partial J$ \cite{ST}.

We hope that the present results trigger further nontrivial developments in nonequilibrium physics.


We wish to thank Glenn Paquette for crucial discussions which made us realize a flaw in the earlier version of the present work.
We also thank the anonymous referee for letting us know of \cite{Ruelle}.
This work was supported by  grants  Nos.  18740240 (TK), 19540392 (NN), and 19540394 (SS)  from the Ministry of Education, Science, Sports and Culture of Japan.



\begin{thebibliography}{10}
\bibitem{en:average}
To be precise, we define expectations with various initial and final conditions as follows.
We omit $\patha$ for simplicity.
$\sbkt{f}_{\Gi,\Gf}
=\{\rhost_{\af}(\Gf)\}^{-1}\int_{\Gamma(0)=\Gi,\Gamma(\Ttot)=\Gf}\calD\pathG\,\WT[\pathG]\,f(\pathG)$,
$\sbkt{f}_{\Gi,\ste}
=\int_{\Gamma(0)=\Gi}\calD\pathG\,\WT[\pathG]\,f(\pathG)$,
$\sbkt{f}_{\ste,\Gf}
=\{\rhost_{\af}(\Gf)\}^{-1}\int_{\Gamma(\Ttot)=\Gf}\calD\pathG\,\rhost_{\ai}(\Gi)\,\WT[\pathG]\,f(\pathG)$,
and
$\sbkt{f}_{\ste,\ste}
=\int\calD\pathG\,\rhost_{\ai}(\Gi)\,\WT[\pathG]\,f(\pathG)$.
Here the normalization factors are simplified by assuming that 
$\rhost_{\af}(\Gf)=\int_{\Gamma(0)=\Gi,\Gamma(\Ttot)=\Gf}\calD\pathG\,\WT[\pathG]$ for any $\Gi$.
\bibitem{en:step}
An infinitesimal protocol $\patha$ is defined by $\alpha(t)=\ai$ for $t\in[0,\Ttot/2)$ and $\alpha(t)=\af$ for $t\in[\Ttot/2,\Ttot]$, where $\af-\ai$ is infinitesimal.
\bibitem{en:Delta}
Let $\Di\beta$ be the typical change in the inverse temperature, and $W$ be the typical work done according to the change of $\pa$.
Then $\Dep\sim E_0\,\Di\beta+\beta\,W$, where $E_0=J_\mathrm{max}\,\tau_\mathrm{r}$  (see \cite{en:Phi})  is a typical energy scale.
\bibitem{en:SD}
Although not all heat baths act as a ``heat source'', 
one can design various source (without violating the 
basic assumption \eqref{e:symmetry}) which keeps the 
averaged  current  almost constant when the parameters 
are changed. An example is a bath with a very high 
temperature coupled very weakly to the system \cite{next}
\bibitem{en:Phi}
We assume that the typical current in the system is $J\sim \kappa\,\delta\beta$ when there is a difference $\delta\beta$ in the inverse temperatures of the baths.
Since the average of $\Phi$ is vanishing, we examine its behavior when the system is disturbed into a non-typical state.
Then one expects a large current of the order $J_\mathrm{max}\sim\kappa\,\beta$.
Since this decays rapidly within the relaxation time $\tau_\mathrm{r}$, we see $\Phi\sim\delta\beta\,J_\mathrm{max}\,\tau_\mathrm{r}\sim \beta\,J_\mathrm{ss}\,\tau_\mathrm{r}\sim\epsilon$, where $J_\mathrm{ss}\sim\kappa\,\delta\beta$ is the current in the steady state.
\end{thebibliography}
\end{document}